\documentclass[conference]{IEEEtran}

\usepackage[pdftex]{graphicx}

\usepackage{array, booktabs}

\newcolumntype{L}[1]{>{\raggedright\arraybackslash}m{#1}} 

\usepackage{balance}
\usepackage{xspace}

\usepackage[pdftex,colorlinks=true,pdfstartview=FitV,
 linkcolor=black,citecolor=black,urlcolor=black]{hyperref}
\usepackage{needspace}

\usepackage{ifthen}
\usepackage[normalem]{ulem} 
\usepackage{xcolor}

\newboolean{showedits}
\setboolean{showedits}{true} 
\ifthenelse{\boolean{showedits}}
{
	\newcommand{\del}[1]{\textcolor{red}{\sout{#1}}} 
	\newcommand{\nbe}[3]{
		{\colorbox{#3}{\bfseries\sffamily\scriptsize\textcolor{white}{#1}}}
		{\textcolor{#3}{\sf\small$\blacktriangleright$\textit{#2}$\blacktriangleleft$}}}
}{
	\newcommand{\del}[1]{} 
	
	\newcommand{\nbe}[3]{}
}
\usepackage[most]{tcolorbox}
\ifthenelse{\boolean{showedits}}
{
  \newtcolorbox{inserted}{%
       title=Inserted text:,
       colframe=blue,colback=blue!5!white,
       breakable,
       leftrule=0mm, 
       bottomrule=0mm,
       rightrule=0mm,
       toprule=0mm,
       arc=0mm, outer arc=0mm,
       oversize
  }
  \newtcolorbox{deleted}{%
       title=Deleted text:,
       colframe=red,colback=red!5!white,
       breakable,
       leftrule=0mm, 
       bottomrule=0mm,
       rightrule=0mm,
       toprule=0mm,
       arc=0mm, outer arc=0mm,
       oversize
  }
  \newtcolorbox{refactored}{%
       title=Rewritten text:,
       colframe=blue,colback=red!5!white,
       breakable,
       leftrule=0mm, 
       bottomrule=0mm,
       rightrule=0mm,
       toprule=0mm,
       arc=0mm, outer arc=0mm,
       oversize
  }
}{

}
\usepackage{amssymb}
\newboolean{showcomments}
\setboolean{showcomments}{false}
\newcommand{\id}[1]{$-$Id: scgPaper.tex 32478 2010-04-29 09:11:32Z oscar $-$}

\ifthenelse{\boolean{showcomments}}
{\newcommand{\nbc}[3]{
 {\colorbox{#3}{\bfseries\sffamily\scriptsize\textcolor{white}{#1}}}
 {\textcolor{#3}{\sf\small$\blacktriangleright$\textit{#2}$\blacktriangleleft$}}}
 }
{\newcommand{\nbc}[3]{}
 }

\newcommand\mg[1]{\nbc{MG}{#1}{brown}}

\newcommand{\ie}{\emph{i.e.},\xspace}
\newcommand{\eg}{\emph{e.g.},\xspace}
\newcommand{\etal}{\emph{et al.}\xspace}
\newcommand{\etc}{\emph{etc.}\xspace}
\newcommand{\smell}[1]{\textbf{#1}\\}
\newcommand{\issue}{\emph{Issue:}\xspace}
\newcommand{\cons}{\emph{Consequently},\xspace}
\newcommand{\symptom}{\emph{Symptom:}\xspace}
\newcommand{\mitig}{\emph{Mitigation:}\xspace}

\renewcommand\sout{\bgroup\markoverwith
	{\textcolor{red}{\rule[0.55ex]{2pt}{0.8pt}}}\color{red}\ULon}

\begin{document}

\title{Relaxing Secure App Development}
\title{Towards Security By Design}
\title{Supporting Android Developers In Making Secure Programming Choices}
\title{Security Smells in Android}

\author{
	\IEEEauthorblockN{Mohammad Ghafari, Pascal Gadient, Oscar Nierstrasz}
	\IEEEauthorblockA{Software Composition Group, University of Bern\\Bern, Switzerland
	}
}

\IEEEoverridecommandlockouts
\IEEEpubid{\makebox[\columnwidth]{\textbf{Preprint -- SCAM 2017}\hfill} \hspace{\columnsep}\makebox[\columnwidth]{ }}

\makeatletter                  
\def\mdseries@tt{m}      
\makeatother                   

\maketitle

\begin{abstract}
The ubiquity of smartphones, and their very broad capabilities and usage, make the security of these devices tremendously important. Unfortunately, despite all progress in security and privacy mechanisms, vulnerabilities continue to proliferate.

Research has shown that many vulnerabilities are due to insecure programming practices. However, each study has often dealt with a specific issue, making the results less actionable for practitioners.

To promote secure programming practices, we have reviewed related research, and identified avoidable vulnerabilities in Android-run devices and the \emph{security code smells} that indicate their presence. In particular, we explain the vulnerabilities, their corresponding smells, and we discuss how they could be eliminated or mitigated during development. 
Moreover, we develop a lightweight static analysis tool and discuss the extent to which it successfully detects several vulnerabilities in about 46\,000 apps hosted by the official Android market.

\end{abstract}

\section{Introduction}

Smartphones and tablets have recently overtaken the number of computers.
They provide powerful features once offered only by computers, 
but the risk of vulnerability on these devices is not on a par with traditional desktop programs; smartphones are increasingly used for security sensitive services like e-commerce, e-banking, and personal healthcare, which make these multi-purpose devices an irresistible target of attack for criminals.

The recent survey on the Stackoverflow website shows that about 65\% of mobile developers work with Android.\footnote{\url{http://insights.stackoverflow.com/survey/2017}}
This platform has captured over 80\% of the smartphone market,\footnote{\url{http://www.gartner.com}}
and just its official app store contains more than 2.8 million apps.
As a result,  a security mistake in an in-house app may jeopardize the security and privacy\footnote{In short, referred to as security in this paper} of billions of users.


The security of smartphones has been studied from various perspectives such as the device manufacturer~\cite{Wu:2013}, its platform~\cite{Xu:2016}, and end users~\cite{Jones:2015}.
Manifold security APIs, protocols, guidelines, and tools are proposed.
%
%
Nevertheless,
security concerns, in effect, are outweighed by other concerns~\cite{Balebako:2014}. 
Many developers undermine their significant role in providing security~\cite{Xie:2011}.
%
As a result, 
apps still suffer from serious proliferating security issues.\footnote{\url{http://www.cvedetails.com}}
For instance, the analysis of 100 popular apps downloaded at least 10M times, revealed that over 90\% of them, due to development mistakes, are prone to SSL vulnerabilities that allow criminals to access credit card numbers, chat messages, contact list, files, and credentials~\cite{Onwuzurike:2015}.


Given these premises, the primary goal of this work is to shed light on the root causes of programming choices that compromise users' security. In contrast to previous research that has often dealt with a specific issue, we study this phenomenon from a broad perspective.
We introduce the notion of \emph{\textbf{security code smells}} \ie \emph{symptoms in the code that signal the prospect of a security vulnerability}.
We have identified avoidable vulnerabilities, their corresponding smells in the code; and discuss how they could be eliminated or mitigated during development. We have also developed a lightweight static analysis tool to look for several of the identified security smells in 46\,000 apps.
In particular, we answer the following three research questions:

\begin{itemize}

\item \textbf{RQ$_{1}$}: What are the security code smells in Android apps?
We have reviewed major related work, especially those appearing in top-tier conferences/journals, and identified 28 avoidable vulnerabilities and the smells that indicate their presence.
We thoroughly discuss each smell, the risk associated with it, and its mitigation during app development.

\item \textbf{RQ$_{2}$}: How prevalent are security smells in benign apps?
We have developed a lightweight tool that statically analyzes apps for the existence of ten security smells. We applied the tool to a repository of about 46\,000 apps hosted by Google.
We realized that despite the diversity of apps in popularity, size, and release date, the majority suffer from at least three different security smells.

\item \textbf{RQ$_{3}$}: 
To which extent identifying security smells facilitates detecting vulnerabilities?
We manually inspected 160 apps, and compared our findings to the result of the tool. Our investigation showed that the identified smells are in fact a good indicator of security vulnerabilities.

\end{itemize}
To summarise, this work represents an initial effort to spread awareness about the impact of programming choices in making secure apps. 
We argue that this helps developers who develop security mechanisms to identify frequent problems, and also provides developers inexperienced in security with caveats about the prospect of security issues in their code.

The remainder of this paper is structured as follows. 
We present the identified vulnerabilities and corresponding security smells in section~\ref{vulAndSmells}. We study the prevalence of these smells and discuss the results in section~\ref{empiricalStudy}. We summarize the work closely related to this paper in section~\ref{relatedWork}, and we conclude the paper in section~\ref{conclusion}.






\section{Security Smells} \label{vulAndSmells}

Although Android security is a fairly new field, it is very active, so researchers in this area have published a large number of articles in the past few years.
We were essentially interested in any paper explaining an issue, or a countermeasure that involves the security of apps in Android.
We used a keyword search over the title and abstract of papers in IEEE Xplore and ACM Digital Library, as well as those indexed by the Google Scholar search engine.  
We formulated a search query comprising \emph{Android} and any other security-related keywords such as 
\emph{security}, \emph{privacy}, \emph{vulnerability}, \emph{attack}, \emph{exploit}, \emph{breach}, \emph{leak}, \emph{threat}, \emph{risk}, \emph{compromise}, \emph{malicious}, \emph{adversary}, \emph{defence}, or \emph{protect}.
We read the title and, if necessary, skimmed the abstract of each paper and included security-related ones.
We further read the introduction of these papers and excluded those whose concerns were not about app security.
In order to extend the search, for each included paper we also recursively looked at both citations and cited papers.
Finally, we carefully reviewed all remaining papers.
During the whole process, we resolved any disagreement by discussion.

We identified 28 smells that may lead to vulnerabilities in Android-powered devices.\footnote{We define vulnerability as a security issue that compromises user's security and privacy.}
We group these smells into five categories.
We explain each smell, its consequence \ie potential risk, and its symptom \ie an identifiable property in the code. We also mention any possible resolution \ie a more secure practice to eliminate or mitigate the issue during app development.

\subsection{Insufficient Attack Protection}

\begin{itemize}
	\item \smell{Unreliable Information Sources}
	Developers acquire their programming knowledge from various sources such as official documentations, books, crowd sources, \etc
	\issue
	According to recent research, developers increasingly resort to studying code examples provided by informal sources like StackOverflow, which are easy to access and integrate, but often lack security concerns~\cite{Acar:2017a}.
	\cons vulnerabilities could make their way into apps in the absence of security expertise.
	\symptom Existence of copy-pasted code from untrustworthy sources.
	\mitig 
	Use official sources which are more reliable, and vet the security of any external code before and after integration in your code.
	\mg{Sites like Stack Overflow are valuable in helping programmers sort out problems they know they have, but don’t help programmers with problems they don't know they may have; most security problems are likely to be of this second type.}
\end{itemize}

\begin{itemize}
	\item \smell{Untrustworthy Libraries}
	Developers cope with the complexity of modern software systems and speed up the development process by relying on the functionalities provided by off-the-shelf libraries.
	\issue 
	Many third-party libraries are unsafe by design \ie introduce vulnerabilities and compromise user data~\cite{Watanabe:2017}. 
	\cons the ramification of adopting such libraries could be manifold. 
	\symptom
	The app utilises unsafe libraries such as advertising libraries that are known to be prone to data leakage~\cite{Soteris:2016}.
	\mitig 
	Solely use reliable libraries~\cite{Backes:2016a}.
\end{itemize}

\begin{itemize}
	\item \smell{Outdated Library}
	The risk of using third-party libraries is not resolved by only using trusted libraries per se.
	\issue 
	Libraries usually offer various bug fixes and improvements in newer releases, but 
	often different developers maintain libraries and apps, and their update cycles generally do not coincide.
	\cons a security breach in an old library or a deprecated API could lead to serious issues. 
	\symptom An included library is behind the latest release, or the app exercises a deprecated API that is not maintained anymore  (\eg the \texttt{SHA1} cryptographic hash function).
	\mitig Integrate the latest release of a library into your app and replace deprecated APIs with their newer counterparts.
	Publish an update not only when the app itself has some improvements but also when there is a new version of a library which the app uses.
\end{itemize}

\begin{itemize}
	\item \smell{Native Code}
	Developers often incorporate native code in their apps to perform intensive computations or to use many third-party libraries which exist in this form.
	\issue Native code is hard to analyze; there is no distinction between code and data at the native level, and attackers can load and execute code from native executables, in a variety of ways much easier than in Java.
	\cons 
	native code is susceptible to severe vulnerabilities like buffer overflow, and
	an attacker could exploit such vulnerabilities, for instance, to execute malicious code~\cite{Wang:2013a}.
	\symptom Existence of native code or a native code library in the app.
	\mitig Use native code only when necessary, and only integrate trustworthy libraries~\cite{Backes:2016a} into your code.
\end{itemize}

\begin{itemize}
	\item \smell{Open to Piggybacking}
	Android apps are often easy to repackage.
	\issue 
	Adversaries could add their malicious code to a benign app before repackaging it~\cite{Li:2017a}. 
	\cons depending on the original app's popularity, users can be infected when installing a seemingly benign app that has evaded the analyses of leading app markets~\cite{Chen:2015}.
	\symptom No technique (\eg watermarking, signature checking) is applied to hardening repackaging.
	\mitig
	Leverage obfuscation to make retro-engineering of apps harder. Also, verify the app's authenticity before any sensitive operation.
\end{itemize}

\begin{itemize}
	\item \smell{Unnecessary Permissions}
	The use of protected features on Android devices requires explicit permissions, and developers occasionally ask for
	more permissions than necessary~\cite{Taylor:2016}.
	\issue 
	The more permission-protected features an app can access, the more sensitive data it can reach. 
	\cons 	
	a more permission-hungry app may expose users to additional security risks~\cite{Taylor:2017}.
	\symptom 
	The manifest file contains permissions for APIs that are not used.
	\mitig Utilize tools like PScout\footnote{\url{http://pscout.csl.toronto.edu}} to exclude from the manifest file any permission whose corresponding API calls are absent in the app.
\end{itemize}



\subsection{Security Invalidation}

\begin{itemize}
	\item \smell{Weak Crypto Algorithm}
	The fundamental set of cryptograph algorithms can be categorized into symmetric, asymmetric, and hash functions.
	\issue
	Each category includes several algorithms, each of which may have various features and attack resilience.
	\cons
	incautious adoption of an algorithm could subject to security issues.
	\symptom
	The use of weak cryptographic hash functions like \texttt{SHA1} or \texttt{MD5}, insecure modes \eg \texttt{ECB} for block ciphers.
	\mitig Consult the state of the art guidelines to choose an appropriate cryptography, and utilize expert systems~\cite{Arzt:2015}.
\end{itemize}

\begin{itemize}
	\item \smell{Weak Crypto Configuration}
	The majority of security breaches come from exploiting developer's mistakes.
	\issue
	Cryptography APIs are widely perceived as being complex with many confusing options~\cite{Nadi:2016}.
	\cons
	a strong but poorly configured algorithm could jeopardise the in-place security.
	\symptom
	Each algorithm has different parameters, and cryptographic parameters in each library could have different defaults. 
	PBE (password-based encryption) with fewer than 1000 iterations, 
	short keys and salts, or none random seeds and initialisation vectors are common mistakes.
	\mg{Check weak certificate and other symptoms in the excel file}
	\mitig
	Use libraries that provide strong documentation and working code examples, and rely on simplified APIs with secure defaults~\cite{Acar:2017b}.
\end{itemize}

\begin{itemize}
	\item \smell{Unpinned Certificate}
	Digital certificates are needed to ensure secure communication.
	Unpinned certificates are easy to maintain and are frequently used in the appified world~\cite{Oltrogge:2015}.
	\issue
	Ensuring the authenticity of a certificate is non-trivial, if it is not pinned.
	\cons
	an app may inadvertently end up trusting a certificate issued by an adversary who has intercepted network communication.
	\symptom
	The app uses unpinned certificates.
	\mitig
	Pinning certificates are always recommended to increase the security.\footnote{Since Android 6.0 pinning can be enabled using the  \texttt{Network Security Configuration} feature.}
	\mg{Hash the certificate to ensure it matches to the intended server.}
\end{itemize}

\begin{itemize}
	\item \smell{Improper Certificate Validation}
	Android provides a built-in process for validating the certificates signed by the trusted Certificate Authorities (CA). 
	\issue 
	In other cases, \eg when a certiifcate  is self-signed, the OS devolves this validation process to the app itself. However, developers often fail to implement it properly~\cite{Fahl:2012}. \cons this leaves the communication channel over SSL/TLS insecure and susceptible to man-in-the-middle attacks~\cite{Conti:2016}.
	\symptom
	The presence of a \texttt{X509TrustManager} or a \texttt{HostNameVerifier} that does not perform any validity check.
	The \texttt{TrustManager} may only use \texttt{checkValidity} to assess the expiration of a certificate without any further check, \eg verifying the certificate's signature or asking the user consent to trust a self-signed certificate.
	Overridden \texttt{onReceiveSslError} in WebView which ignores any certification errors.
	\mitig  
	Ensure the certificate chain is valid \ie 
	the root certificate of the chain is issued by a trusted authority,
	none of the certificates in the chain are expired, and 
	each certificate in the chain is signed by its immediate successor in the chain. 
	Moreover, the certificate should match its designated destination, \ie the ``Common Name" field or the ``Subject Alternative Name" in the certificate should match the domain name of the server being connected to.
	Finally, utilize network security testing tools like ``Nogotofail"\footnote{\url{https://github.com/google/nogotofail}} to examine your communication.
\end{itemize}

\begin{itemize}
	\item \smell{Unacknowledged Distribution}
	Google Play, Google's official marketplace for Android, strives to identify potential security enhancements when an app is uploaded to it.
	However, developers may distribute their packages via other channels to circumvent out-of-order updates, bypassing the slow release cycles and security restrictions of this market place.
	\issue The protection provided by Google, including code and signature checks, is neglected. 	
	\cons the risk of distributing a vulnerable app increases especially when the app utilises uncertified libraries, or in a worse case, an attacker can replace installation packages with malicious ones \cite{Zheng:2014}.
	\symptom The \texttt{android.permission.INSTALL{\_}PACKAGES} permission exists in the manifest.
	\mitig Distribute your apps and updates exclusively through official app stores that perform security checks.
\end{itemize}

\subsection{Broken Access Control}

\begin{itemize}
	\item \smell{Unauthorised Intent Receipt} 
	An \emph{intent} is an abstract specification of an operation that apps can use to utilise the actions provided by other apps.
	An \emph{explicit} intent guarantees communication with the specified recipient, but it is the Android system that determines the recipient(s) of an \emph{implicit} intent among available apps.
	\issue
	Any app that declares itself able to serve the requested operation is potentially eligible to fulfill the intent.
	\cons 
	if such an app is malicious, a threat called \emph{intent hijacking} could arise in which user information carried by the intent could be manipulated or leaked~\cite{Chin:2011}.
	\symptom
	The existence of an intent with private data, but without a particular component name (the fully-qualified class name).
	\mitig
	Only use explicit intents for sending sensitive data. In addition, always validate the results returned from other components to ensure they comply with your expectation.
\end{itemize}

\begin{itemize}
	\item \smell{Unconstrained Inter-Component Communication}
	One app can reuse components (e.g., activities, services, content provider, and broadcast receivers) of other apps, provided those apps permit it.
	\issue 
	Android apps are independently restricted in accessing resources.
	\cons 
	a threat called \emph{component hijacking} arises when a malicious app escalates its privilege for originally prohibited operations through other apps that access those operations~\cite{Wu:2016,Davi:2010}.
	\symptom The existence of the \texttt{intent-filter} element or \texttt{android:exported = true} attribute in the manifest file without any permission check to ensure that a client app is originally permitted to receive that service.
	\mg{Up to API level 19 if the export attribute misses \texttt{android} prefix, a component is per default exported}
	\mitig Exclusively export components that are meant to be accessed from other apps
	and avoid placing any critical state changing actions within such components. Enforce custom permissions with the \texttt{android:permission} attribute to prohibit access from apps with lower privileges.
	Finally, use tools like \emph{IccTA}, which detects flaws in inter-component communication~\cite{Li:2015}.
\end{itemize}

\begin{itemize}
	\item \smell{Unprotected Unix Domain Socket}
	Android IPCs do not support cross-layer IPC, \ie communication between an app's Java and native processes/threads. To circumvent this limitation developers resort to using Unix domain sockets. Moreover, developers may reuse Linux code that already utilizes such sockets.
	\issue Developers are barely guided to protect Unix domain sockets with appropriate authentication.
	\cons adversaries are capable of abusing these exposed IPC channels to exploit vulnerabilities within privileged system daemons and the kernel~\cite{Shao:2016}. 
	\symptom
	The server socket channel accepts clients without performing any authentication or similarly a client connects to a server without properly authenticating the server.
	\mitig
	Enforce proper security checks when using the sockets.
\end{itemize}

\begin{itemize}
	\item \smell{Exposed adb-level Capabilities}
	Android Debug Bridge (adb) is a versatile tool that provides communication with a connected Android device. 	Many developers opt for adb-level capabilities to legitimately access a subset of signature-level resources~\cite{Lin:2014}.
	\issue 
	For this purpose, an app communicates locally with an adb-level proxy through the TCP sockets opened on the same device, which exposes the adb server to any app with the INTERNET permission.
	\cons a malign app with ordinary permissions can command the adb and establish serious attacks~\cite{Hwang:2015}.
	\symptom The existence of adb-specific commands or TCP connection to local host in the code.
	\mitig Avoid using adb-level capabilities in your app, as it is also prohibited since Android 6.0.
\end{itemize}

\begin{itemize}
	\item \smell{Debuggable Release}
	During app development there exist two major build configurations, debug and release. The first is meant for active development, while the latter is for signed in-market releases.
	However, developers may forget to switch to release mode before publishing an app~\cite{Xu:2013}. 
	\issue Apps shipped with debugging enabled always try to connect to a local Unix socket opened by the Android Debug Bridge (adb). While adb is not running on every consumer device, a malign app could disguise itself as an adb service and connect to random debuggable apps.
	\cons a malicious app is able to gain full access to the Java process and can execute arbitrary code in the context of the debuggable app \cite{MWRInfoSecurity:2011}.
	\symptom The manifest file contains the attribute \texttt{android:debuggable = true}.
	\mitig The debug mode should be disabled in the signed release version \ie either the debuggable attribute should not exist in the manifest file, or its value should be false. More recent build environments already perform this task automatically. 
\end{itemize}

\begin{itemize}
	\item \smell{Custom Scheme Channel}
	Scheme channels (a.k.a protocol prefixes) like \texttt{fblite://} for Facebook allow seamless interactions between web and Android apps.
	\issue The sender of a scheme message is not able to verify the recipient of the message so that malign apps could register themselves as a receiver of another app's unified resource identifier (URI) scheme.
	\cons adversaries could collect access tokens or other sensitive information~\cite{Wang:2013b}. 
	\symptom The registration of a URI scheme within the \texttt{intent-filter} in manifest file. The \texttt{SchemeRegistry.register} method is in the code.
	\mitig 
	Adopt the dedicated system scheme \ie \texttt{Intent} which is harder to compromise.
\end{itemize}

\subsection{Sensitive Data Exposure}

\begin{itemize}
	\item \smell{Header Attachment}
	The header section of data transport protocols like HTTP comprises key/value pairs to store operational parameters.
	\issue Developers may rely on headers to transfer sensitive data, \eg they store credentials to auto-login into a service. 
	\cons any adversary eavesdropping on the network may easily access the attached data \cite{Wang:2013b}.
	\symptom Calls like \texttt{HttpGet.addHeader()} are present in the code to store private data.
	\mitig Do not store  sensitive data in headers, instead rely on dedicated mechanisms like OAuth2 protocol\footnote{\url{https://oauth.net/2}} to authenticate to third-party services.
\end{itemize}

\begin{itemize}
	\item \smell{Unique Hardware Identifier}
	Each device often has a couple of globally unique identifiers such as the IMEI number, MAC address, \etc
	\issue 
	For various purposes like user profiling, apps utilize these IDs, which are tied to each device.
	\cons anyone in the possession of such IDs would be able to track the user's activities across various sources.
	\symptom Method calls that return IDs from associated classes like \texttt{TelephonyManager} or \texttt{BluetoothAdapter} exist in the code.
	\mitig Use the \texttt{UUID.randomUUID()} API to ensure that the retrieved ID is globally unique for each user, but only within the same app identity.
\end{itemize}

\begin{itemize}
	\item \smell{Exposed Clipboard}
	Users usually rely on a clipboard to copy and paste data across apps.
	\issue The clipboard content is readable and writable by all apps.
	\cons a malign app could perform versatile attacks on the clipboard content from URL hijacking to data exfiltration and code injection \cite{Zhang:2014a}.
	\symptom 
	The related calls on \texttt{ClipboardManager} exist in the code.
	The app uses the common \texttt{TextView} and \texttt{EditText} controls, which allow copy and paste to handle sensitive data~\cite{Nan:2017}.
	\mitig 
	Never allow sensitive data to be copied and pasted in your app. Perform input validation before exercising any input from the clipboard.
\end{itemize}

\begin{itemize}
	\item \smell{Exposed Persistent Data}
	Android provides various storage options to store persistent data. 
	These options vary depending on the size, type, and accessibility of data.\footnote{\url{https://developer.android.com/guide/topics/data/data-storage.html}}
	\issue
	Developers may opt for a particular option without considering its security implication.
	\cons they
	expose private data.
	\symptom 
	The existence of a private storage with global access scope (\ie \texttt{MODE\_WORLD\_READABLE} or \texttt{MODE\_WORLD\_WRITEABLE}) in the app. The app relies on \texttt{ContentProvider} to access data, but there is no access restriction for other apps.
	\mg{default export value for content provider is true in Android 16 or below. later the default has changed to false
	The DOM storage or FileAccess is enabled in \texttt{WebView}.}
	%
	\mitig
	Specify permissions to protect who can access your shared data.  
	Encrypt any (internally or esp. externally) stored sensitive data, and place the encryption key in KeyStore, protected with a user password that is not stored on the device.
	%
\end{itemize}

\begin{itemize}
	\item \smell{Insecure Network Protocol}
	Data transportation channels exist in various flavours, and insecure ones like HTTP are more prevalent and easy to maintain.
	\issue 
	Insecure channels transfer data without encryption per se.
	\cons an attacker can secretly relay the data and possibly alter it~\cite{Onwuzurike:2015}.
	\symptom
	APIs related to opening insecure network connections like \texttt{http} or \texttt{ftp} exist in the code\mg{if the server only accepts https this won't be a problem - Developers may disable TLS functionality during testing and never re-enable it.}.
	\mitig All app traffic should happen over a secure channel. Otherwise, any sensitive data should be encrypted before it is sent out. Android 6.0 or above provides the \texttt{cleartextTrafficPermitted} property which protects app from any usage of cleartext traffic.
\end{itemize}

\begin{itemize}
	\item \smell{Exposed Credentials}
	Passwords, private keys, secret keys, certificates, and other similar credentials are commonly used for authentication, communication, or data encryption. 
	\issue In some circumstances such data is inadvertently disclosed to unauthorised parties.
	\cons this could break the intended security.
	\symptom
	The app contains hard-coded credentials, or they are stored without any password protection such as when the \texttt{KeyStore.ProtectionParameter} is null. \mg{There are more symptoms}
	\mitig
	Store such data in a KeyStore in a protected format which restricts unauthorised accesses.
\end{itemize}

\begin{itemize}
	\item \smell{Data Residue}
	According to recent research, about 80\% of abandoned apps are likely to be uninstalled in less than a week~\cite{Liu:2017}.	
	\issue 
	After an app is uninstalled, various types of data associated to the app, ranging from its permissions, operation history, configuration choices, and so on may still remain in a few system services~\cite{Zhang:2016c}.
	\cons 
	such so-called ``data residue" can be associated to another app and empower adversaries to access sensitive information~\cite{Zhang:2016b, Zhang:2016c}.
	\symptom
	The app calls system services that are known to be subject to data residue problem.
	%
	\mitig
	Unfortunately, an app may not always be aware of its data being stored in system services, 
	and the mere mitigation is to avoid sharing private data with these services, if possible.
\end{itemize}

\subsection{Lax Input Validation} 

\begin{itemize}
	\item \smell{XSS-like Code Injection}
	WebView is an essential component that enables developers to use web technologies such as HTML and JavaScript to deliver 	web content within an app. Unlike Web browsers like Chrome, FireFox, \etc which are developed by well-recognized companies that we trust, each app using a WebView is like a customized browser which may not have undergone thorough security tests.
\issue An app may load web content unsafely \ie without sanitising the input from any code. 
\cons an adversary could inject malicious code through any channel that the app uses to get web content~\cite{Jin:2014}.
\symptom
The \texttt{setJavaScriptEnabled} call with value \texttt{true} which enables execution of JavaScript exists in the code, and the app fetches web content from untrustworthy sources (\eg by calling \texttt{loadUrl} or \texttt{loadData} on \texttt{WebView}) without applying proper sanity checks.
\mitig
Invoke the default browser to display untrusted data. Use a HTML sanitizer to filter out any code inside the data, and show plain text only using safe APIs that are immune to code injection (\ie do not execute JavaScript code).
Beware of third-party libraries that employ WebView. Disable JavaScript, if you do not need it.

\end{itemize}

\begin{itemize}
	\item \smell{Broken WebView's Sandbox}	
	There is a sandbox inside WebView that separates its JavaScript from the rest of system.
	\issue 
	WebView provides an API, \texttt{addJavascriptInterface}, through which an app can 
	access Java APIs, and therefore mobile resources, from within JavaScript code inside the sandbox.
	\cons if the app renders the web content unsafely, a  code injection attack is possible~\cite{Jin:2014}.
	\symptom
	In addition to the symptoms of the previous issue, the \texttt{addJavascriptInterface} call exists in the code.	
	\mitig
	Take into account the suggestions of the previous issue, and as well
use the \texttt{@JavascriptInterface} annotation to specify any method that is exposed by JavaScript to prevent reflection-based attacks.

\end{itemize}

\begin{itemize}
	\item \smell{Dynamic Code Loading}
	Android allows apps to load and execute external code and resources.
	\issue 
	Although dynamic code loading is widely adopted, developers are often unaware of the risks associated to this generally unsafe mechanism or fail to implement it securely~\cite{Poeplau14}.
	An attacker can replace the code that is to be loaded with a malicious one.
	\cons this can lead to severe vulnerabilities such as remote code injection~\cite{Falsina:2015}.
	\symptom Use of any class loader in the code.
	In case of loading the code and resources of another installed app, a call to \texttt{createPackageContext()} on the \texttt{Context} object exists in the code. 
	\mitig 
	Either bundle the required resources within each app package, or verify the integrity and authenticity of the loaded code \eg by imposing restrictions on its location or provenance~\cite{Titze:2015}.
	Analyze your app with the help of tools like \emph{Grab 'n Run}~\cite{Falsina:2015}.
\end{itemize}

\begin{itemize}
	\item \smell{SQL Injection}
	Data-driven apps organize their data through a database.
	\issue
	An app might directly use inputs to build a query that will be run by the database engine.
	\cons
	an adversary who succeeds at inserting malicious code into SQL statements, can access or modify database data.
	\symptom
	Inputs from untrustworthy sources are passed to the database without proper validation.
	\mitig
	Instead of dynamic SQL generation, rely on parameterized queries and stored procedures which let the database distinguish between code and data. Validate inputs and filter suspicious values \eg \emph{escape characters} to ensure they do not end up in the query.
	\mg{Untrustworthy sources are end user, intents, clipboard, \etc}
\end{itemize}


\mg{WebView.postUrl causes leaky URLs}

\section{Empirical study} \label{empiricalStudy}

We developed a lightweight analysis tool that statically detects known security smells in an app. We rely on the Apktool to reverse engineer Android apk files and generate smali code.\footnote{\url{https://ibotpeaches.github.io/Apktool}}
\mg{about smali: https://github.com/JesusFreke/smali}
We defined a set of rules to capture the symptoms of each security smell. In particular, we utilize Java XML Parser for parsing the Manifest files and use regular expressions to define and match the code pattern corresponding to the identified symptoms of each smell in the code.

We randomly selected our apps from the AndroZoo dataset.\footnote{\url{https://androzoo.uni.lu}}
This dataset currently provides more than 5.5M apps collected from several sources.
We initially collected a random subset of 70\,000 apps whose sources are in Google Play. 
However, to collect more meta data information such as an app's category, its number of downloads, update cycle, and star rating we still needed to visit the Google Play website. 
Unfortunately, we could not access 25\,000 apps for various reasons, for example, because they were no longer available on the store, or they were not accessible from Switzerland. In the end, we included 46\,000 benign apps in our dataset.
About 90\% of these apps were released between 2014 and 2016,
a quarter of them were updated within three months,
the majority
were rated more than four stars,
slightly more than 27\%
were downloaded above 50\,000 times,
and the median apk size was 5.5MB.

\subsection{Result}
\begin{figure}
\includegraphics[width=\columnwidth, trim=2cm 3cm 2cm 2cm]{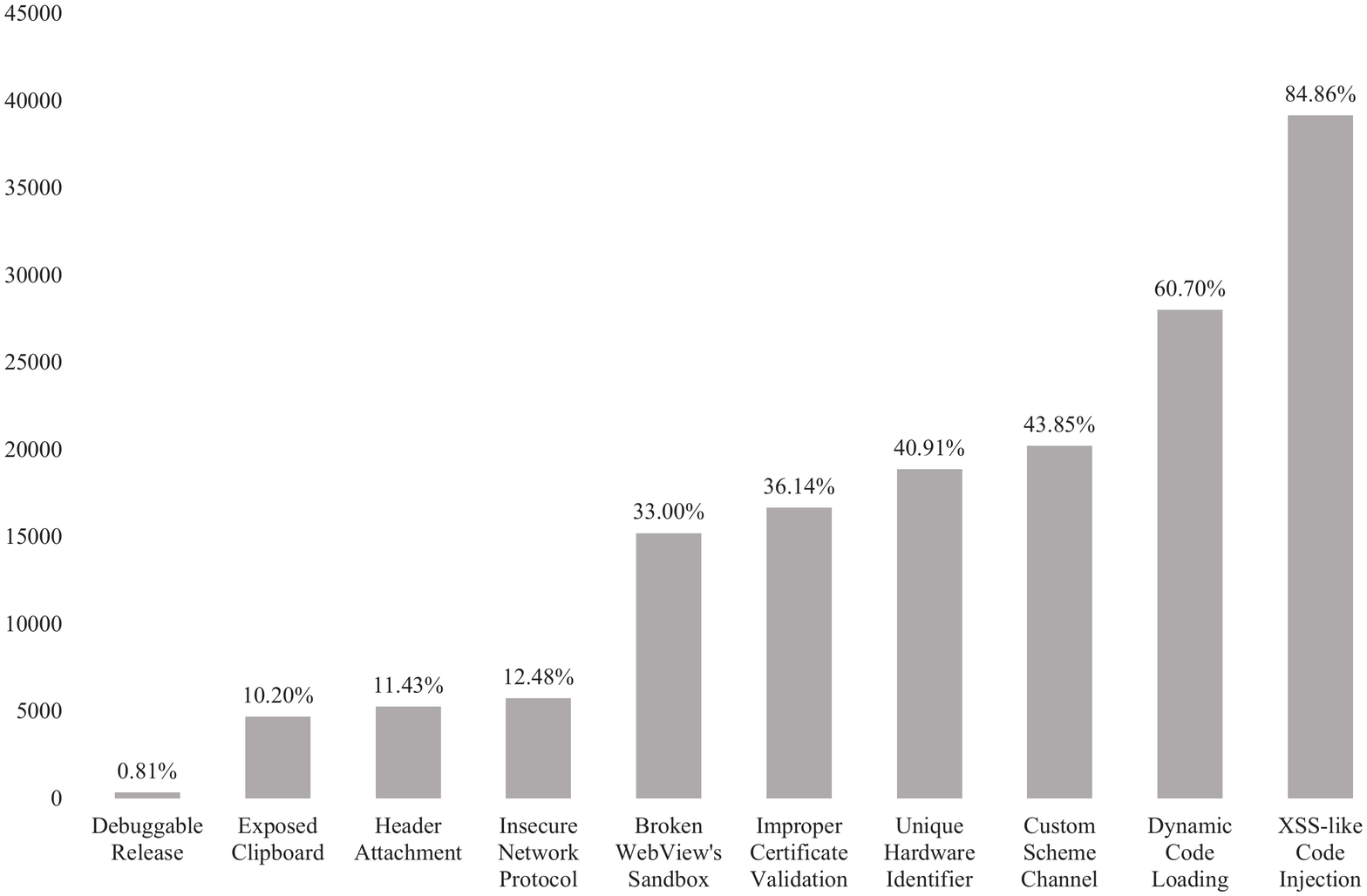}
\caption{Distribution of security smells in the apps}
\label{fig:apps_per_issue}
\end{figure}

We applied our lightweight tool to all apps in the dataset.
Figure~\ref{fig:apps_per_issue} shows how prevalent the smells are in our dataset. 
A majority of apps potentially suffer from XSS-like code injection (85\%) followed by dynamic code loading (61\%). About 40\% use custom scheme channels and expose a unique hardware identifier. More than 12\% use an insecure network protocol, and almost 10\% are subject to header attachment as well as clipboard issues.
Finally, just under 1\% of the apps have debug mode enabled.

\begin{figure}
\includegraphics[width=\columnwidth, trim=2cm 4cm 2cm 3cm]{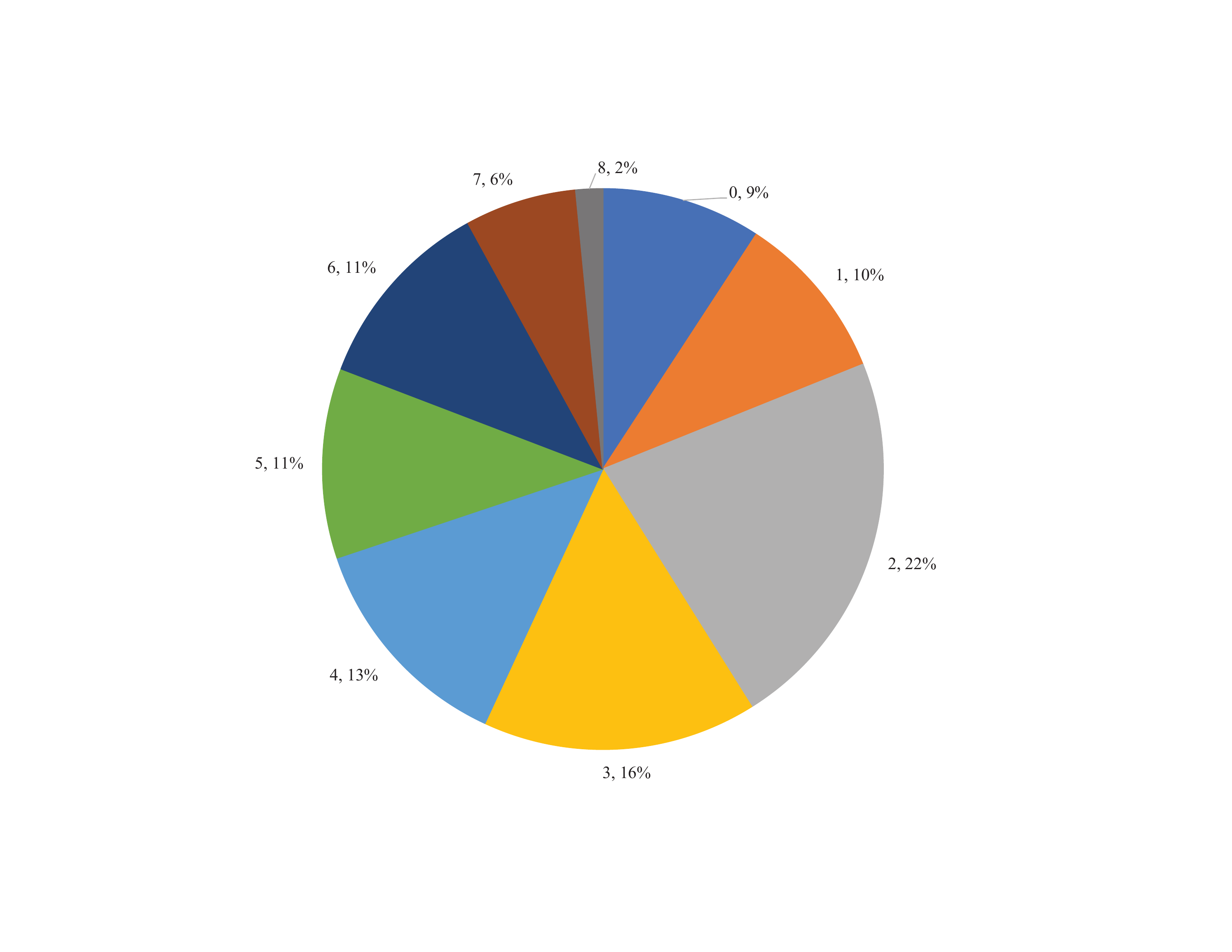}
\caption{Partitioning apps by number of security smells}
\label{fig:apps_grouped_by_number_of_issues}
\end{figure}

We also studied how many of security smells usually appear in the apps (see Figure~\ref{fig:apps_grouped_by_number_of_issues}).
Only 9\% of apps are free of any smell, a majority \ie above 50\% suffer from at least three different smells, and over a quarter are subject to more than four smells, which is catastrophic.

\begin{figure}
\includegraphics[width=\columnwidth, trim=2cm 3cm 2cm 2cm]{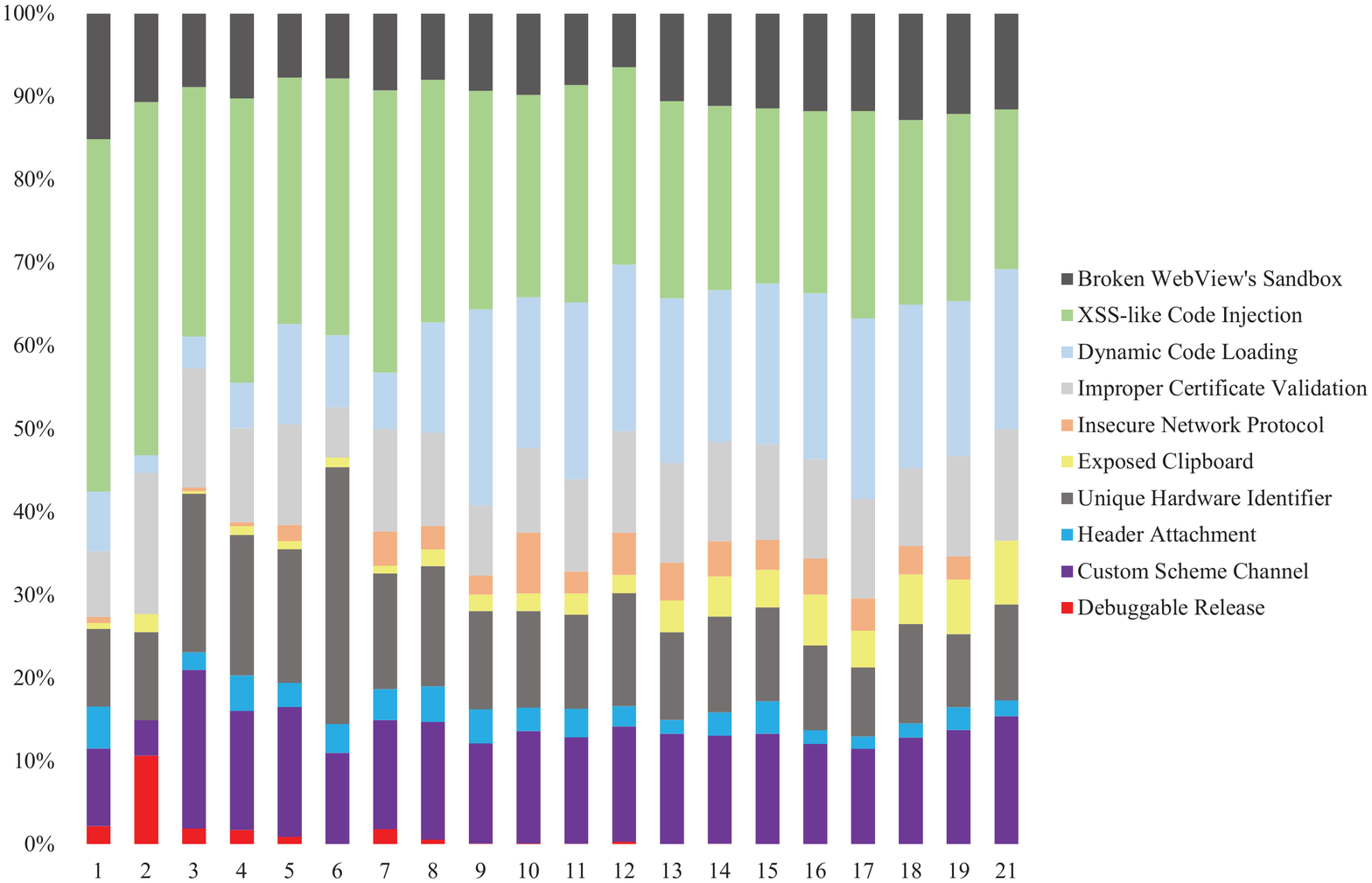}
\caption{The distribution of security smells within each API level}
\label{issuesPercentageAPI}
\end{figure}

We also investigated the prevalence of security smells at different API levels as the proportion of devices running different API versions varies. Figure~\ref{issuesPercentageAPI} shows the distribution of smells within each API level. 
We noticed that the prevalence of \emph{Debuggable Release} has been dramatically reduced. We believe this is mainly due to the fact that Google market no longer accepts apps in debug mode. We conjecture this issue should have decreased also in other markets without this constraint as recent build platforms automatically disable the debug mode in the signed release version.
In contrast, there is an increase in the existence of the \emph{Exposed Clipboard} security smell. This could stem from the many sharing options for social media in the apps. 
Similarly, the issue of \emph{Dynamic Code Loading} has become more common. We observed that many developers adopt this feature to implement their own update mechanisms.

\begin{figure}
\includegraphics[width=\columnwidth, trim=2cm 3cm 2cm 2cm]{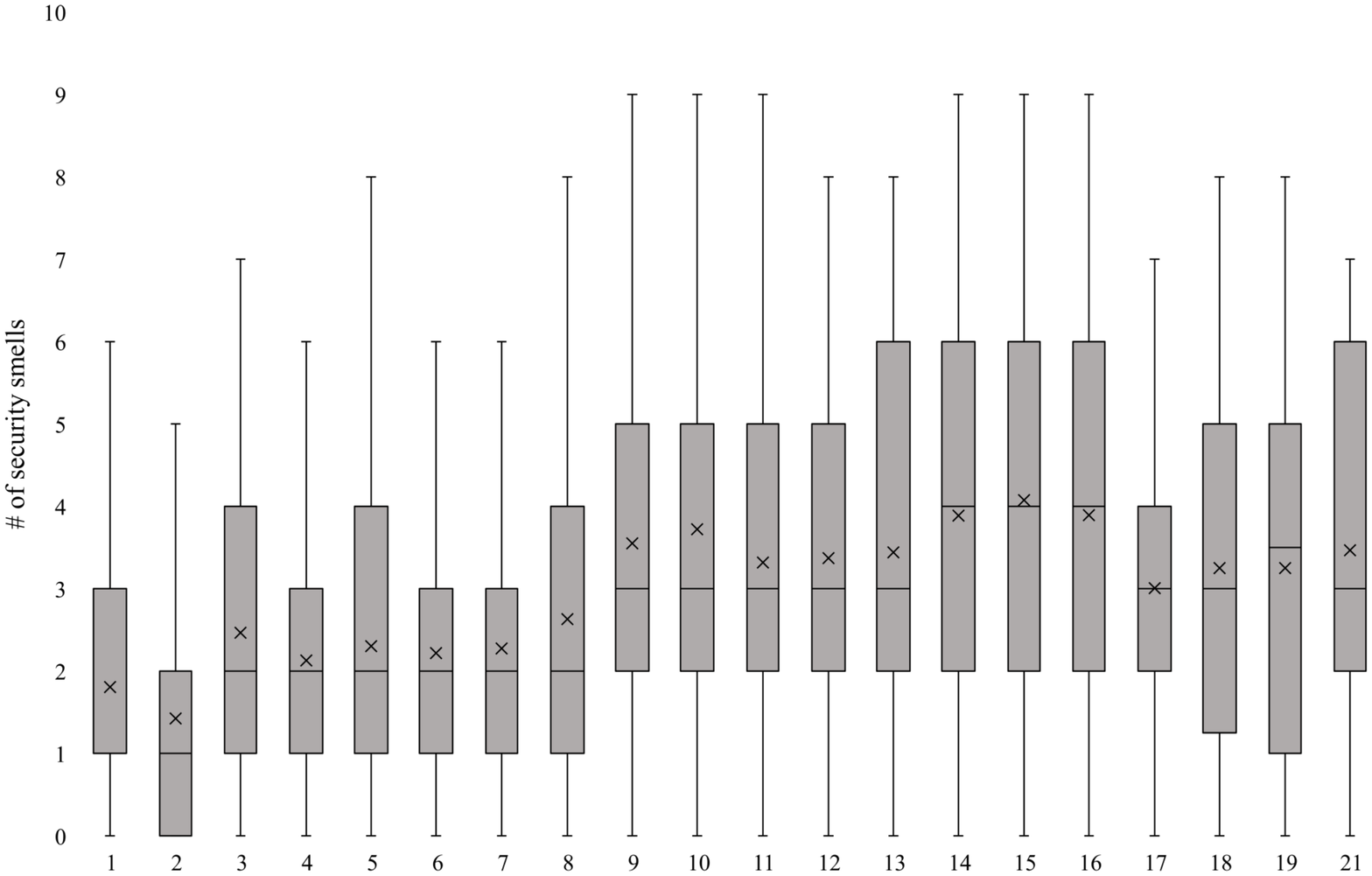}
\caption{Number of smells within each API level}
\label{issuesPerAPI}
\end{figure}

Figure~\ref{issuesPerAPI} shows how many of these classes of smells appear within each API level.
There is a correlation between feature availability and feature usage, and apparently these uses have introduced more insecurity. It seems the peak of issues was reached around API level 15.
%

In the remainder of this section we discuss our findings from a few more perspectives.
\mg{justify opting for distinct issues}

\subsubsection*{Category}
\begin{figure*}
\includegraphics[width=\textwidth, trim=2cm 3cm 2cm 2cm]{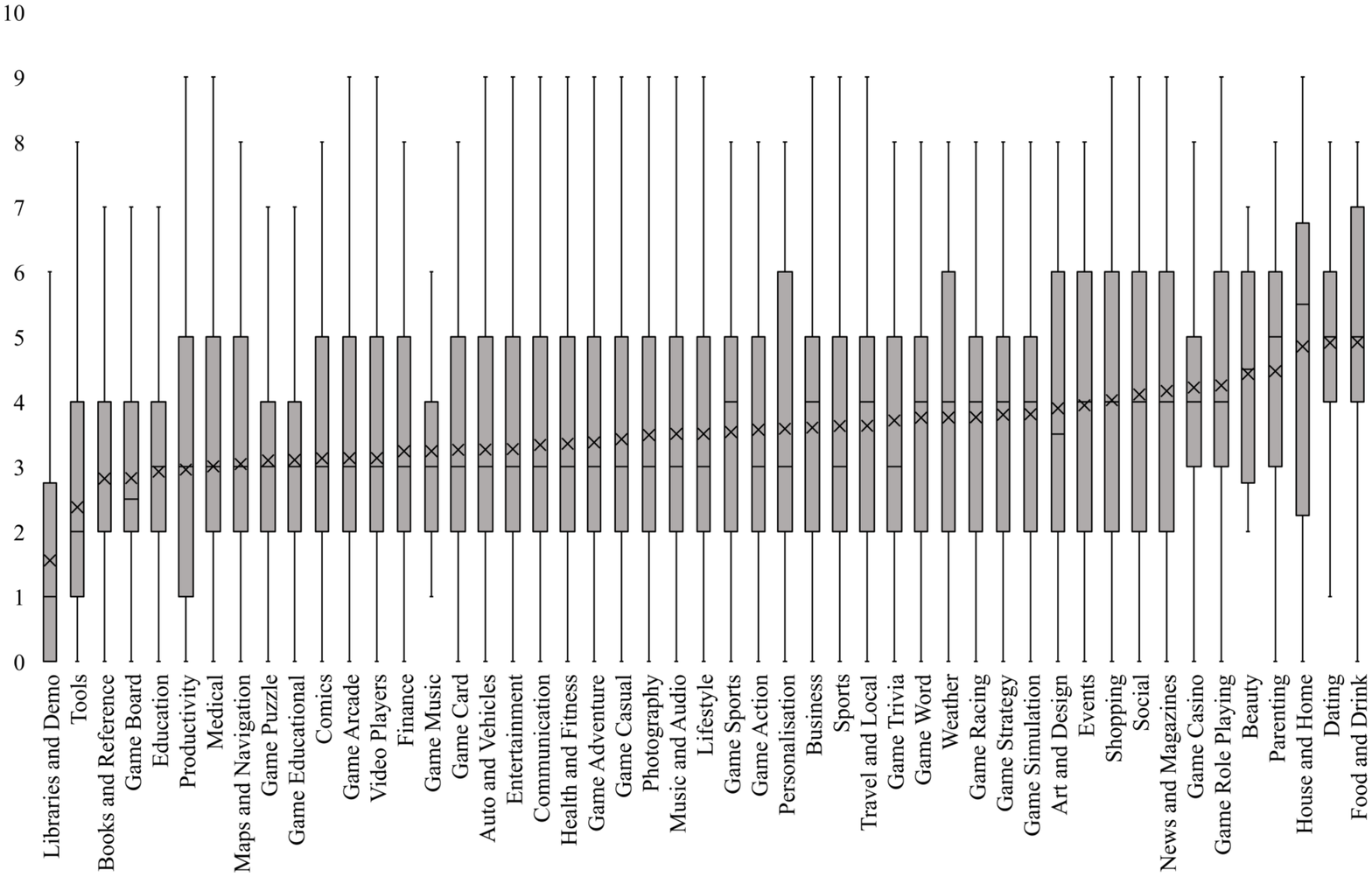}
\caption{Distribution of smells in app categories}
\label{prevalenceOfDistinctIssuesInAppCategories}
\end{figure*}

Figure~\ref{prevalenceOfDistinctIssuesInAppCategories} shows the number of different security smells appearing in the apps in each category.
The apps in the \emph{Libraries and Demo} category are the most secure ones as they usually rely on local content. 
We noted that security smells are prevalent in gaming apps, and that \emph{Casino} and \emph{Role Playing} games are more problematic.
Finally, \emph{Dating} as well \emph{Food and Drink} apps suffer from the highest number of security smells.

\subsubsection*{Popularity}

\begin{figure}
\includegraphics[width=\columnwidth, trim=2cm 2cm 2cm 2cm]{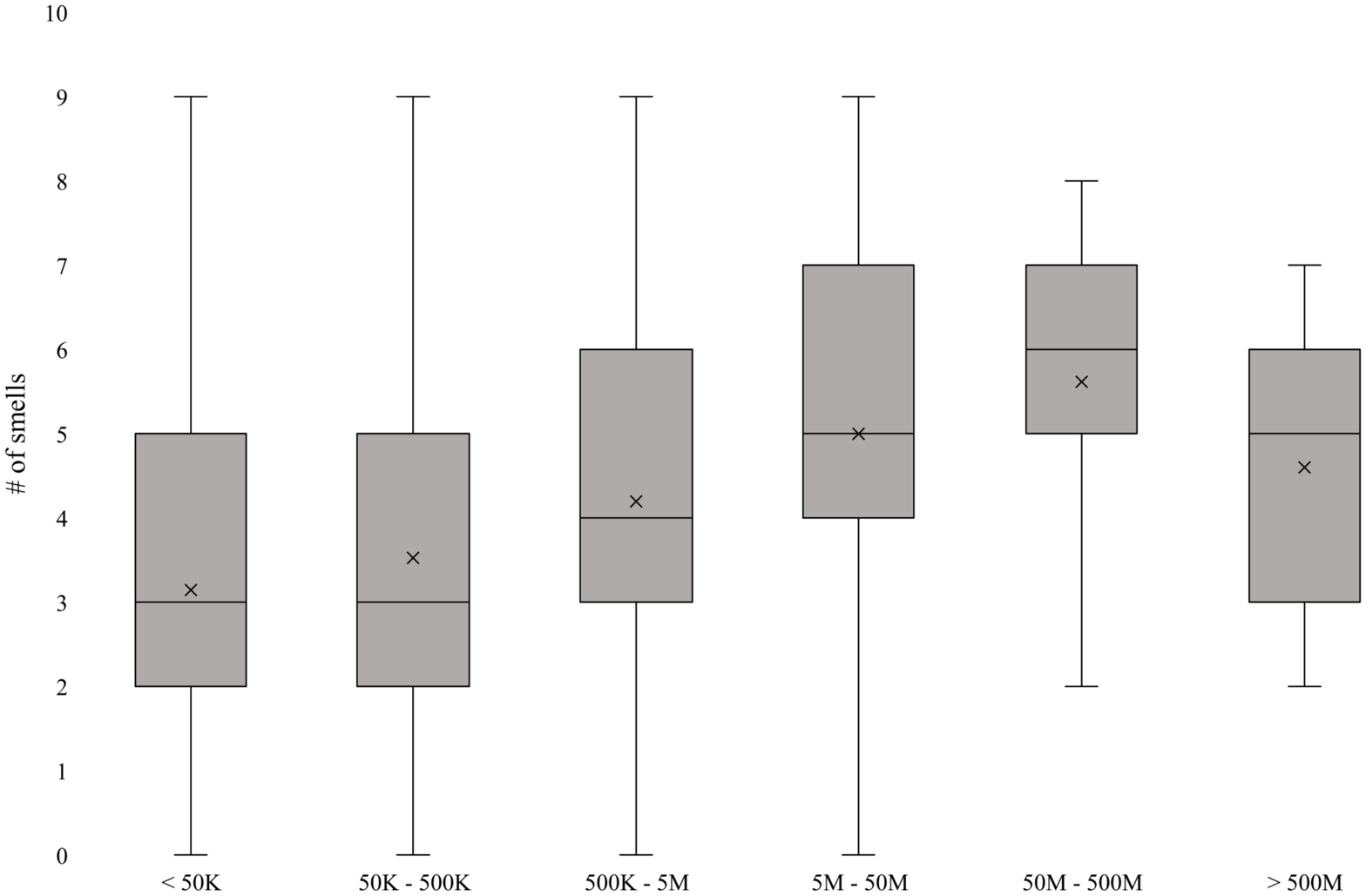}
\caption{The relationship between number of smells and number of downloads}
\label{issueDownload}
\end{figure}

\begin{figure}
\includegraphics[width=\columnwidth, trim=2cm 2cm 2cm 2cm]{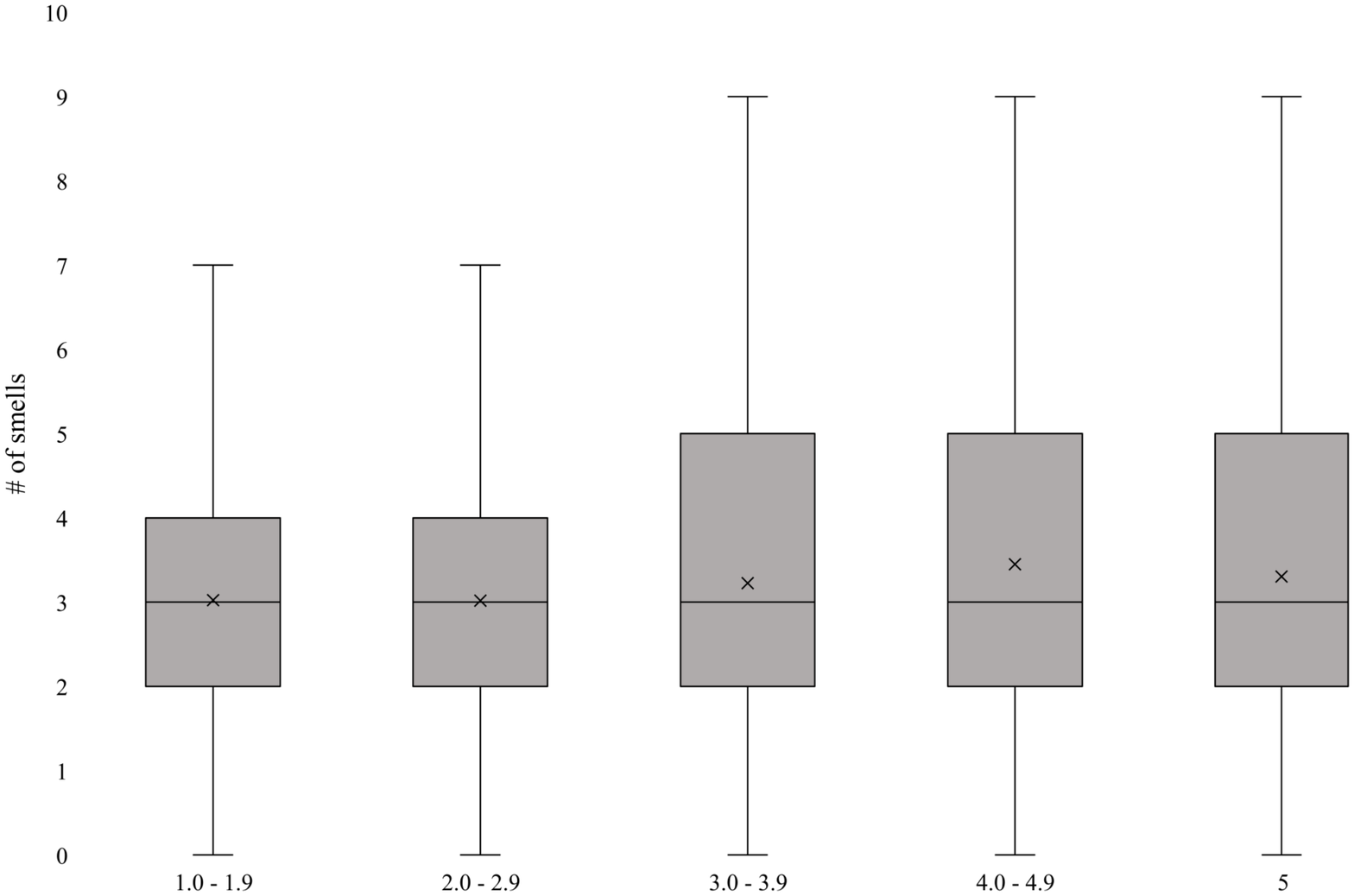}
\caption{The relationship between number of smells and app star ratings}
\label{issueStar}
\end{figure}

Figure~\ref{issueDownload} shows the relationship between the number of downloads and the security smells. The majority of apps with millions of downloads suffer from five kinds of smells.
Although about 73\% of apps within our dataset were downloaded less than 50\,000 times, there were still enough apps with more downloads to conclude that the number of downloads never guarantees security.  
\mg{We further noticed these apps vary in size, and there is no relationship between the number of downloads and the size of apps}
Figure~\ref{issueStar} shows the relationship between the number of security smells and star ratings. Despite the number of stars, apps often suffer from three kinds of security smells. In particular, the star rating correlates negatively with the presence of security smells. We assume that the studied security smells are barely noticeable by end-users, hence they are not reflected in the ratings.

\subsubsection*{Release date}

We further studied whether the prevalence of security smells changes over time. In fact, with advances in developers support (\eg tools, learning resources) we expected that security smells in more recent apps should be rarer than in older apps. Nonetheless, the result showed that neither the number of smells nor the likelihood of a particular smell relates to the release date of an app. 
Moreover, we noted that in general the security of apps with short update cycles is similar to those with longer update cycles.
That is, either security issues in one release still remain in future releases, or they get fixed but new releases also introduce new smells.

\subsubsection*{App Size}

We were interested to know whether existence of security smells is ever related to the size of an app. Our investigation showed that an app may suffer from various kinds of security smells, despite its size. In fact, increase in app size may only increase the frequency of a security smell. It is also worth mentioning that some apps are larger not because of having more code but other resources such as image, video and audio content.

\subsection{Manual Analysis}

\begin{figure}
\includegraphics[width=\columnwidth, trim = 2cm 2cm 2cm 2cm]{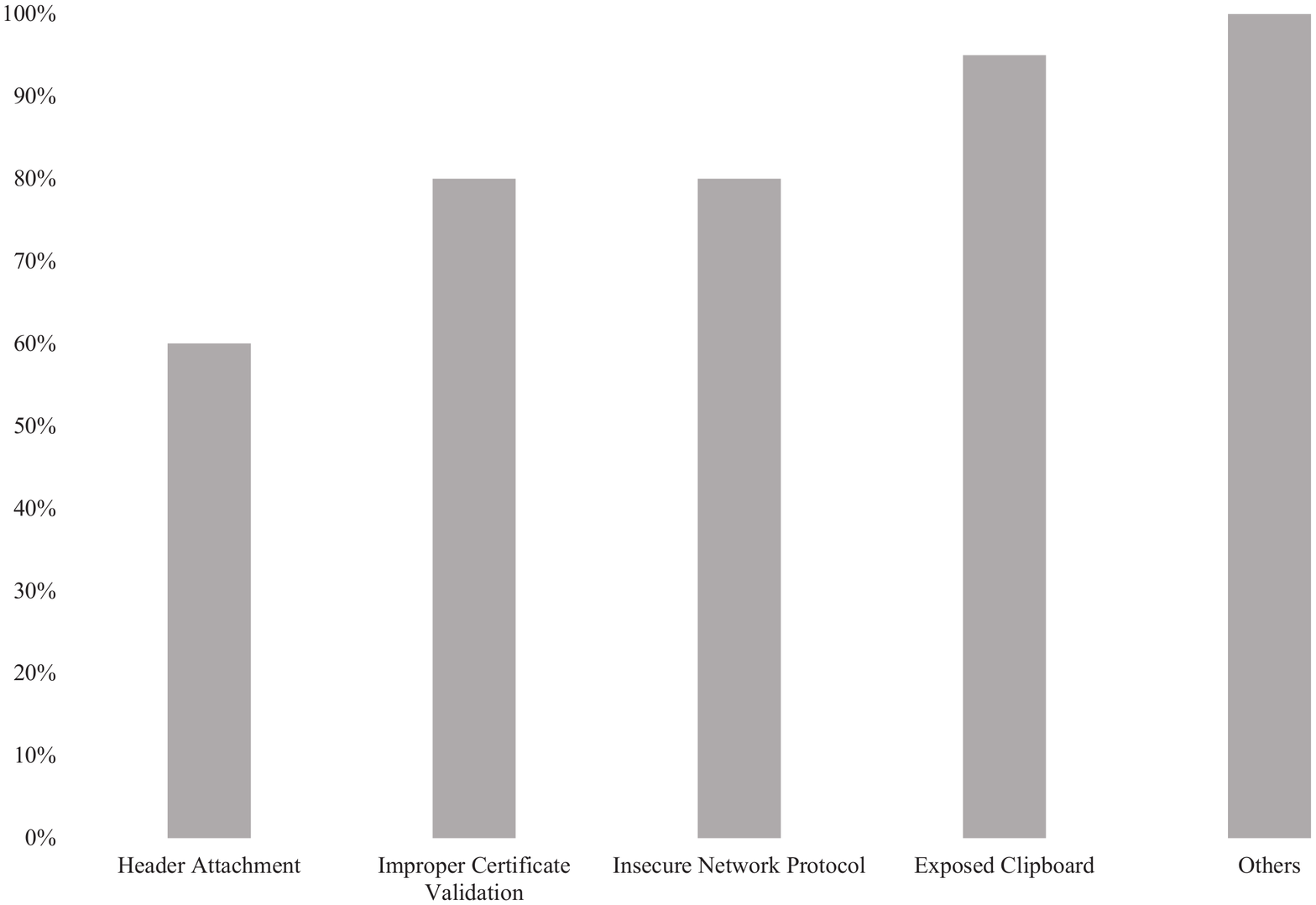}
\caption{The precision of obtained results}
\label{manualAnalysis}
\end{figure}

To assess how reliable these findings are to detect security vulnerabilities, we manually analyzed 160 apps. For each smell, we inspected 20 apps manually and compared our findings to the result of the lightweight analysis tool.
\mg{We miss evaluating false negatives!}
As is shown in Figure~\ref{manualAnalysis}, the results were encouraging. 
The manual analysis completely agreed with the tool in the security risk associated with six security smells. 
In case of exposed clipboard the tool achieved a very good performance \ie above 90\% agreement with the manual analysis.
The level of agreement in insecure network protocol and improper certificate validation was 80\%. 
We realized some apps use http connections to exclusively load local contents which is legitimate in  development frameworks like Apache Cordova or Adobe PhoneGap. And some apps implemented their own custom \texttt{TrustManager} which in fact was secure.
Finally, our tool was unable to correctly detect the security risk associated with header attachment in 40\% of cases, which is mainly due to the fact that discerning data sensitivity is non-trivial.

\subsection{Threats to Validity}
We note several limitations and threats to validity of the results pertinent to our research.
One important threat is the completeness of this study \ie whether we could identify and study all related papers in the literature. 
We could not review all the publications, but we strived to explore top-tier software engineering and security journals and conferences as well as highly cited work in the field.
For each relevant paper we also recursively looked at both citations and cited papers. 
Moreover,
to ensure that we did not miss any important paper, for each identified issue we further constructed more specific queries and looked for any new paper on GoogleScholar.

We analyzed the existence of security smells in the source code of an app, whereas third-party libraries could also introduce smells.

We were only interested in studying benign apps as in malicious ones developers may not spend any effort to accommodate security. Thus, we merely collected apps which were available on official Google market. However, our dataset may still have malicious apps that evaded the security checks of the market.

Finally, the fact that the results of our lightweight analysis tool are validated against manual analysis performed by the authors is a threat to construct validity through potential bias in experimenter expectancy.

\section{Related Work}
\label{relatedWork}

%
Reaves~\etal discussed Android specific challenges to program analysis and assessed android application analysis tools, and found that they mainly suffer from lack of maintenance, and are often unable to produce functional output for applications with known vulnerabilities~\cite{Reaves:2016}.
Li~\etal studied the state-of-the-art work that statically analyses Android apps~\cite{Li:2017b}. They found that much of this work supports detection of private data leaks and vulnerabilities, a moderate amount of research is dedicated to permission checking, and only three studies deal with cryptography issues. Unfortunately, much state-of-the-art work does not publicly share their artefacts.
Linares-Vasquez~\etal mine 660 Android vulnerabilities available in the official Android bulletins and the CVE-details and present a taxonomy of the their types; they report the presence of those vulnerabilities affecting the Android OS and acknowledge that most of them can be avoided by relying on secure coding practices~\cite{Linares-Vasquez:2017}.
Finally, Sadeghi~\etal review 300 research papers related to Android security, and provide a taxonomy to classify and characterize the state-of-the-art research in this area~\cite{Sadeghi:2016}. They find that 26\% of existing research is dedicated to vulnerability detection, but each study is usually concerned with specific types of security vulnerabilities.
%
%
%
Our work expands on such studies to provide practitioners with an overview of the security issues that are inherent in insecure programming choices.

Some research is devoted to educating developers for secure programming.
Xie~\etal interviewed 15 professional developers about their software security knowledge, and realized that many of them have reasonable knowledge but do not adopt it as they believe it is others' responsibility~\cite{Xie:2011}.
Weir~\etal performed open-ended interviews with a dozen app security experts, and identified that app developers should learn analysis, communication, dialectic, feedback and upgrading in the context of security~\cite{Weir:2016b}.
Witschey~\etal surveyed developers about their reasons for adopting and not adopting security tools~\cite{Witschey:2015}. Interestingly, they found the perceived prestige of security tool users and the frequency of interaction with security experts more important to promote security tool adoption.
Acar~\etal suggest a high-level research agenda to achieve usable security for developers. They propose several research questions to elicit developers' attitudes, needs, and priorities in the area of security~\cite{Acar:2016b}.
Our work is complementary to these studies in the sense that provides an initial assessment of developers' security knowledge, and as well highlights the significant role of developers in making more secure apps.

\section{Conclusion} \label{conclusion}

In contrast to all advances in software security, software systems are suffering from increasing security and privacy issues. Security in Android, the dominant mobile platform, is even more crucial as these devices often contain manifold sensitive data, and a security issue in a small home-brewed app can threaten the security of billions of users.

To fundamentally reduce the attack surface in Android, we promote the adoption of secure programming practices. We reviewed state of the art papers in security and identified 28 smells whose presence may indicate a security issue in an app. We developed a static analysis tool to study the prevalence of ten of such smells in 46\,000 apps.
We realized that despite the diversity of apps in popularity, size, and release date, the majority suffer from at least three different security smells.
Moreover, the manual inspection of 160 apps showed that the identified security smells are actually a good indicator of security vulnerabilities.

\section{Acknowledgments}
We gratefully acknowledge the funding of the Swiss National Science Foundations for the project ``Agile Software Analysis" (SNF project No. 200020\_162352, Jan 1, 2016 - Dec. 30, 2018).
\footnote{\url{http://p3.snf.ch/Project-162352}}

\balance

\bibliographystyle{abbrv}

\bibliography{reference} 

\end{document}